\documentclass[prb,showpacs,superscriptaddress,twocolumn,notitlepage]{revtex4-1}

\usepackage{graphicx, subfigure}

\usepackage{amsmath}
\usepackage{amsfonts}
\usepackage{graphicx}

\usepackage{natbib}
\usepackage[final=true]{hyperref}

\begin{document}
\title{Drag of electrons in graphene by substrate surface polar phonons}

\author{S.~V.~Koniakhin}
\email{kon@mail.ioffe.ru}
\affiliation{Ioffe Physical-Technical Institute of the Russian Academy of Sciences, 194021 St.~Petersburg, Russia}
\affiliation{St. Petersburg Academic University - Nanotechnology Research and Education Centre of the Russian Academy of Sciences, 194021 St. Petersburg, Russia}

\author{A.~V.~Nalitov}
\affiliation{School of Physics and Astronomy, University of Southampton, Southampton SO17 1BJ, United Kingdom}

\begin{abstract}

It is known that electron scattering by surface polar phonons (SPPs) of the substrate reduces their mobility in supported graphene. However, there is no experimental evidence for contribution of drag of electrons by SPP to thermoelectric phenomena in graphene: graphene thermopower exhibits good agreement with Mott's law, which means that the diffusion contribution to the thermopower is dominant in a wide range of carrier densities and temperatures. Here we develop a complete theory of drag of electrons in graphene by SPP. By solving Boltzmann transport equation for electrons scattered by SPPs we derive SPP drag contribution to the thermopower in graphene. Compared to diffusion thermopower, obtained values appear to be one order of magnitude lower for various substrates. This can be explained by low occupation number of the SPPs and short mean free path of such phonons stemming from their small group velocity. We conclude that experiments on thermopower in graphene can be treated within the framework of Mott's law.

\end{abstract}

\maketitle

\section{Introduction}

During the past decade there have been intense studies of mechanical \cite{frank2007mechanical,lee2008measurement}, electronic \cite{RevModPhys.81.109,RevModPhys.83.407,PhysRevLett.98.186806}, optical \cite{PhysRevB.86.115301,Glazov2014101,PhysRevLett.107.276601}, thermal \cite{6235226,nika2009phonon,nika2012two} and magnetic properties of graphene. Significant role among them plays experimental \cite{PhysRevB.80.081413,PhysRevB.83.113403,PhysRevLett.102.096807,PhysRevLett.102.166808,dollfus2015thermoelectric} and theoretical \cite{PhysRevB.79.075417,PhysRevB.80.235415,PhysRevB.82.155410,vaidya:093711,PSSB:PSSB201248302,koniakhin2013phonon,alisultanov2015thermodynamics,alisultanov2014anomalous} investigation of thermoelectricity.
Intriguing thermoelectric phenomena that contributes to thermopower in graphene, as well as in metals \cite{Gurevich46}, carbon nanocomposites \cite{EidVul} and graphite \cite{PhysRevB.21.2462,PhysRevB.28.2157,PhysRevB.34.4298}, is the effect of phonon drag \cite{koniakhin2013phonon}.

Suspending graphene sheets at a distance of hundreds of nanometers from the substrate gives possibility to investigate native properties of graphene. However, gating the graphene devices and controlling carriers density requires close contact between graphene sheet and the substrate. Therefore various aspects of graphene-substrate interaction are of active current research \cite{R.2010,Frank2014440,doi:10.1021/jp8008404,GrapheneIr111}.
Substrate reduces mobility of carriers in graphene due to electron scattering on surface charged impurities \cite{PhysRevLett.98.186806,doi:10.1143/JPSJ.75.074716}, surface corrugations \cite{doi:10.1021/nl070613a,PhysRevLett.100.016602,Katsnelson195} and atomic steps \cite{AtomicSteps}. Molecular dynamics simulations show that van der Walls interaction between graphene \cite{MD_graphene} or nanotubes \cite{PhysRevB.84.165418} and substrate significantly reduces relaxation time of intrinsic phonons.

In recent studies it was shown that scattering by surface polar phonons (SPPs) in substrates like SiO$_2$ and SiC reduce electron mobility in graphene \cite{PhysRevB.77.195415,PhysRevB.81.195442,Chen2008,bhargavi2013scattering} and carbon nanotubes \cite{doi:10.1021/nl8030086,ref1}. However, current experimental data on thermoelectric properties of supported graphene \cite{PhysRevB.80.081413,PhysRevB.83.113403,PhysRevLett.102.096807,PhysRevLett.102.166808} show that behavior of thermopower in graphene coincides with Mott law that describes the diffusion contribution. Therefore is important to elucidate why the contribution from drag by substrate phonons have not been robustly detected yet.

In this paper we consider a monolayer graphene with linear electron dispersion law $\varepsilon_{\mathbf{k}} = \hbar v_F k$, and degenerate electron gas obeying Fermi statistics. Fermi energy is related to the carrier density by $\varepsilon_F = \hbar v_F k_F = \hbar v_F \sqrt{\pi n}$. It is useful to introduce the dimensionless electron wave vector $\tilde{k} = k a_0/\pi$. The tilde is used to denote other quantities, normalized in the same manner.

\section{Theory}

Surface polar phonons of the substrate generate an electric field at significant distances from the substrate (see fig. 1 from ref. \cite{doi:10.1021/nl8030086}). The phonon-induced field penetrates graphene at the surface of the substrate and provides the probability for an electron in graphehe to be scattered by remote substrate phonon. Electron scattering by SPP is not the dominant mechanism of electron mobility reduce in relatively thick semiconductor layers in metal-oxide-semiconductor field-effect transistors, but plays significant role for graphene and carbon nanotubes.

The electron transition rate arising from scattering by SPPs is given by\cite{PhysRevB.81.195442}

\begin{equation} \label{matrixEl}
W_{\mathbf{k}\rightarrow \mathbf{k}+\mathbf{q}}^{SPP} = A_{\mathbf{k},\mathbf{q}} \frac{4\pi^2e^2F^2}{Sq}\exp(-2qz_0),
\end{equation}
where $\mathbf{k}$ and $\mathbf{q}$ are electron and phonon wave vectors respectively, $e$ is the electron charge, $S$ is a surface of graphene sheet and $z_0 \approx 3.5$ \AA\ is the van der Waals distance between graphene sheet and substrate. Multiplier $A_{\mathbf{k},\mathbf{q}} = \frac{1}{2}(1+\cos(\theta_{\mathbf{k}+\mathbf{q}}-\theta_{\mathbf{k}}))$ arises from the chiral nature of carriers in graphene. The electric field magnitude and consequently the scattering rate $W$ overall are defined by Fr\"{o}lich coupling

\begin{equation} \label{Fsqared}
F^2=\frac{\hbar\omega_{ph}}{2\pi}\left( \frac{1}{\varepsilon_{\infty}+\varepsilon_\textit{env}} - \frac{1}{\varepsilon_{0}+\varepsilon_\textit{env}}\right),
\end{equation}
where $\varepsilon_{\infty}$ and $\varepsilon_{0}$ are low- and high-frequency dielectric constants of the substrate and $\varepsilon_\textit{env}$ is the environment dielectric constant. Following ref. \cite{PhysRevB.81.195442} we assume the latter to be equal to 1. As it can be seen from exponential multiplier in \eqref{matrixEl}, $q^{-1}$ is a characteristic distance at which the electric field decays outside the substrate. In table 2 of ref. \cite{PhysRevB.81.195442} the energies of surface optical phonons and values of Fr\"{o}lich coupling strength for various substrate materials are listed. For variety of substrates $\hbar \omega_{ph} \sim 0.1$\,eV and $F^2$ value is close to 0.5\, meV.

The SPP phonon collision integral, entering the Boltzmann transport equation on the electron distribution function $f(\mathbf{\varepsilon_{\mathbf{k}}})$, can be written down as
\begin{widetext}
\begin{multline} \label{4types}
\left(\frac{\partial}{\partial t} f( k) \right)_{ph} = - \frac{2\pi}{\hbar} \frac{S}{4 \pi^2} \int{d \mathbf{q}} \times\\
\left[
W_{k \rightarrow k +  q}N_{ph}(q)f^{(0)}(\varepsilon_{ k}) \left(1 - f^{(0)}(\varepsilon_{ k +  q}) \right)\delta(\varepsilon_{k}-\varepsilon_{ k+q}-\hbar\omega_{ph}) + \right. \\ W_{ k \rightarrow  k -  q}f^{(0)}\left( N_{ph}(q) +1 \right) f^{(0)}(\varepsilon_{ k}) \left(1 - f^{(0)}(\varepsilon_{ k - q}) \right)\delta(\varepsilon_{k}-\varepsilon_{ k-q}+\hbar\omega_{ph})\\
- W_{ k -  q \rightarrow  k} N_{ph}(q) f^{(0)}(\varepsilon_{ k -  q}) \left(1 - f^{(0)}(\varepsilon_{ k }) \right) \delta(\varepsilon_{k}-\varepsilon_{ k-q}+\hbar\omega_{ph})-\\
\left. W_{ k +  q \rightarrow  k} \left( N_{ph}(q) +1 \right) f^{(0)}(\varepsilon_{ k +  q}) \left(1 - f^{(0)}(\varepsilon_{ k})
\right) \delta(\varepsilon_{k}-\varepsilon_{ k+q}-\hbar\omega_{ph}) \right]\,,
\end{multline}
\end{widetext}
where $f^{(0)}(\varepsilon_{ \mathbf{k}})$ is the equilibrium Fermi distribution function.

In presence of temperature gradient phonon distribution function $N_{ph}(q)$ writes as

\begin{equation} \label{Nph}
N_{ph}(\mathbf{q}) = N_{ph}^{(0)}(q) + N_{ph}^{(1)}(\mathbf{q}).
\end{equation}
In the equation above $N_{ph}^{(0)}(q)$ is the Bose equilibrium distribution function for phonons and $N_{ph}^{(1)}(\mathbf{q})$ is a small correction due to the temperature gradient, which can be written in the relaxation time approximation as

\begin{equation} \label{Nph(1)}
N_{ph}^{(1)}(\mathbf{q})=\tau_{ph}(q) \mathbf{v}(\mathbf{q})\vec{\nabla} T
 \frac{\partial N_{ph}^{(0)}(q)}{\partial T} ,
\end{equation}
where $\mathbf{v}(\mathbf{q}) = \nabla_{\mathbf{q}} \omega_{ph}$ is a group velocity of optical phonons. The dispersion law of optical phonons writes as
\begin{equation} \label{WonQ}
\omega_{ph}(q) = \omega_0 (1-\beta \tilde{q}^2),
\end{equation}
where $\tilde{q} = qa_0/\pi$ is the phonon wave vector of normalized to graphene lattice constant $a_0 = 2.46$ \AA. Approximating optical phonons dispersion curves of many semiconductors and dielectrics allows one to estimate $\beta \sim 0.1$.

We assume that the optical phonon lifetime $\tau_{ph}$ is independent on the phonon wave vector $q$. Multiple studies devoted to molecular dynamics simulations \cite{esfarjani2011heat,he2011heat,turney2009predicting,bao2012first} and Raman experiments \cite{anand1996temperature,aku2005long,song2008direct,letcher2007effects,PhysRevB.61.3391,lee2010comparing} indicate optical phonon lifetime of about dozens of picoseconds in various solids, and we set $\tau_{ph}=10$\,ps. The optical phonon group velocity is linear in $q$, therefore $\tau_{ph}v(q)=L_0\tilde{q}$, where $L_0 = 2 \tau_{ph} \omega_{ph} \beta \frac{a_0}{\pi}$. Taking $\tau_{ph} = 10$\,ps, $\hbar \omega_{ph} = 0.1$\,eV and $\beta = 0.1$ yields $L_0 = 25$\,nm.

Electrons in graphene are scattered by phonons with $q$ of the order of $k_F \ll \pi/a_0$ and for actual values of Fermi energy $\hbar \omega_{ph} \beta  \tilde{k}_F^2 \ll k_BT$. Thus the occupation number of phonons does not depend significantly on magnitude of the phonon wave vector $q$ and the quantity
\begin{equation} \label{dNdT}
\frac{\partial N_{ph}^{(0)}(q)}{\partial T} = k_B\frac{\hbar \omega_{ph}e^{\frac{\hbar \omega_{ph}}{k_BT}}}{\left( e^{\frac{\hbar \omega_{ph}}{k_BT}} - 1 \right)^2 \left( k_BT \right)^2}
\end{equation}
is considered below to be constant.

\begin{figure*}[!ht]
\centering     
\subfigure[]{\label{fig1a}\includegraphics[width=0.49\textwidth]{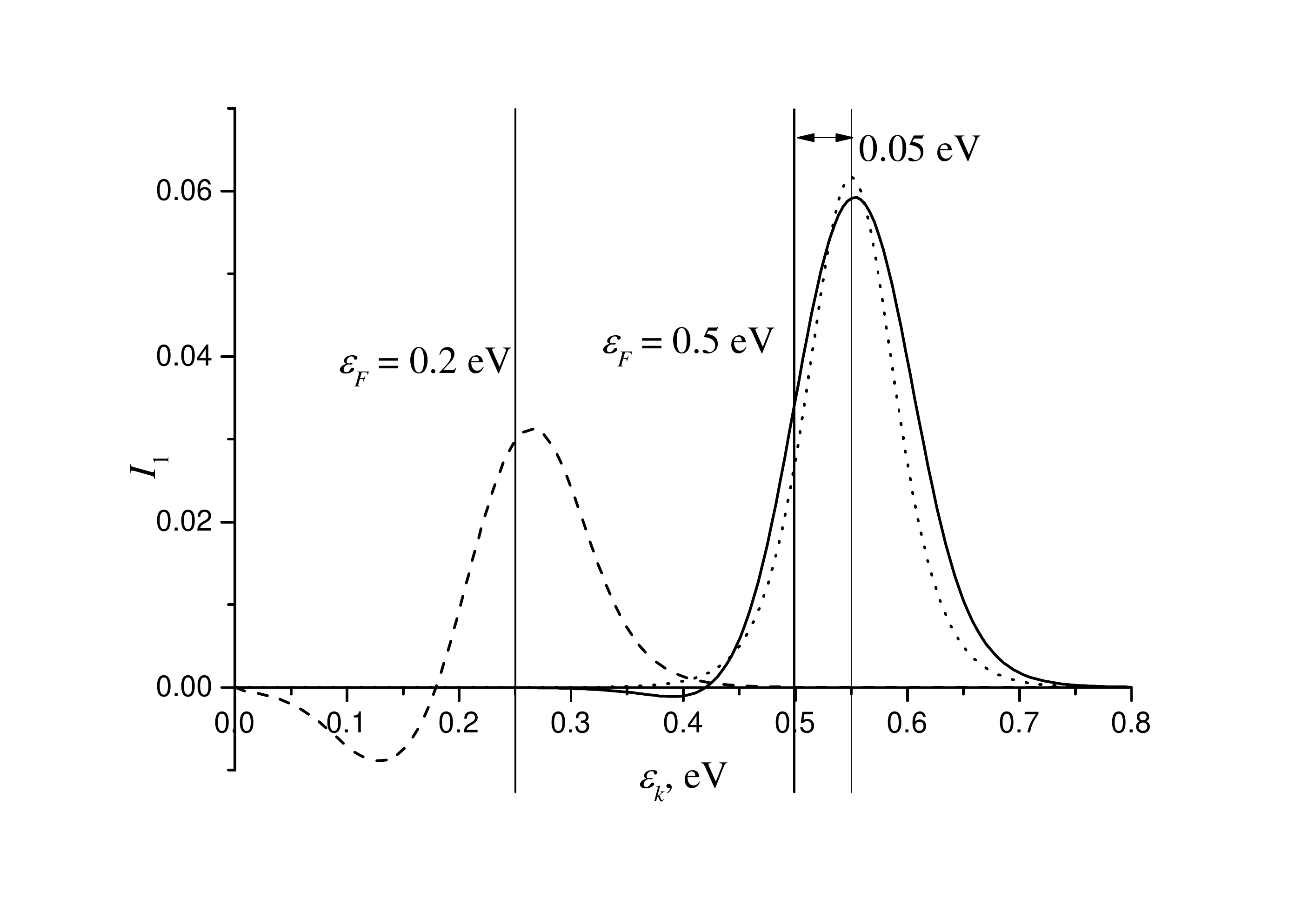}}
\subfigure[]{\label{fig1b}\includegraphics[width=0.49\textwidth]{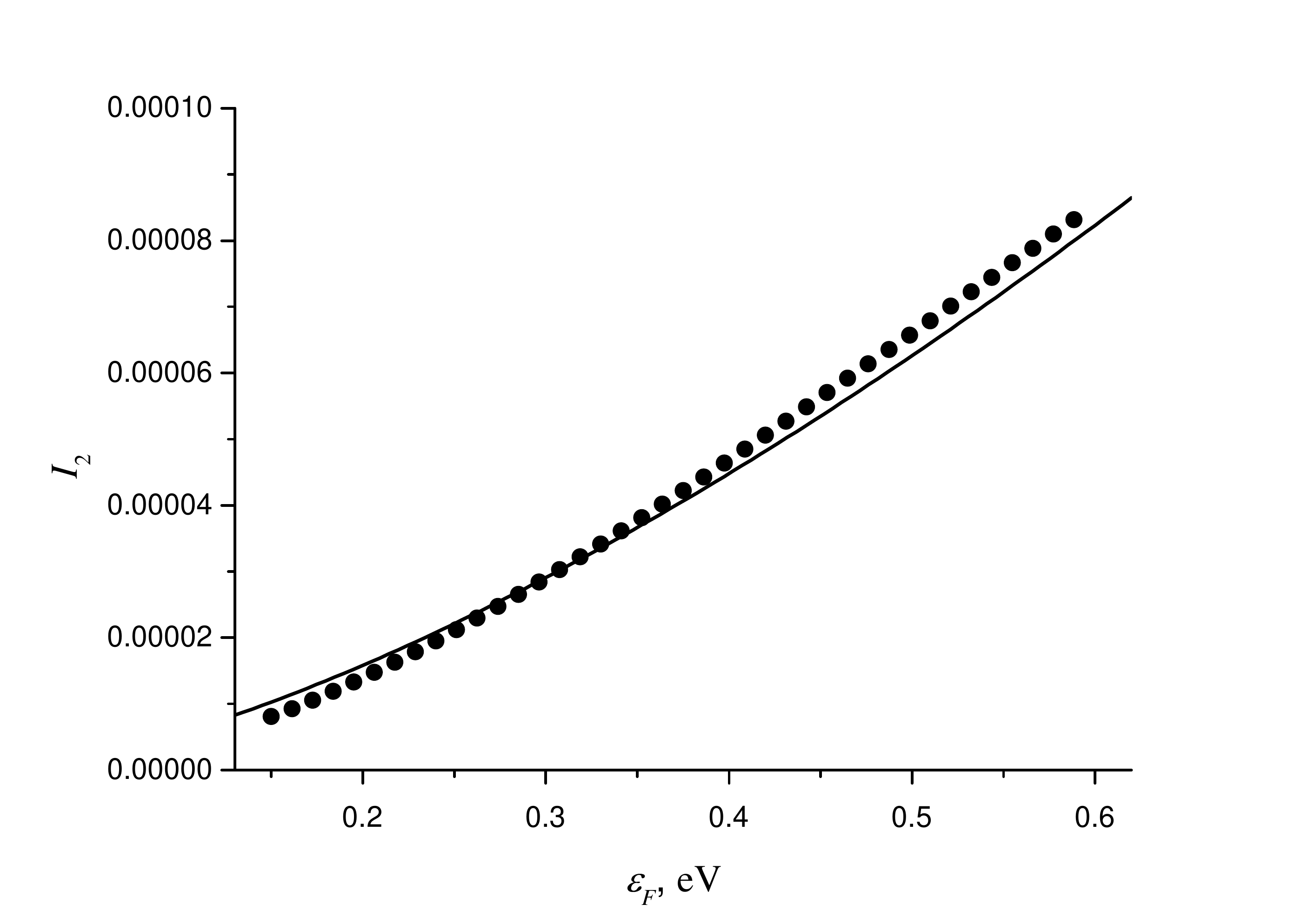}}
\caption{
Panel (a). Dependence of $I_1(\tilde{k})$ magnitude on $\varepsilon_k = \hbar v_F \pi \tilde{k}/a_0$. Optical phonon energy is 0.1\,eV. Solid curve is for $\varepsilon_F=0.5$\,eV and dashed curve is for $\varepsilon_F=0.2$\,eV. Interestingly to notice that for large Fermi energies $I_1(k) \approx \frac{\partial}{\partial \varepsilon}f^{(0)}(\varepsilon-\frac{1}{2}\hbar\omega_{ph})$, denoted by dotted curve. Panel (b). $I_2$ as a function of Fermi energy. $T=300$ K, $\hbar \omega_{ph} = 0.1$ eV. Circles denote results of numerical integration and black line curve shows approximation with \eqref{I2_approx1}.
}
\label{fig1}
\end{figure*}

When $N_{ph}^{(0)}$ is substituted to the right-hand part of \eqref{4types}, the latter vanishes due to energy conservation law entering the delta functions. After replacing $\mathbf{q}$ with $-\mathbf{q}$ in the second and the third terms in the square brackets in \eqref{4types}, substituting first correction for phonon distribution \eqref{Nph(1)}, one obtains the following correction for the electron distribution function within the framework of relaxation time approximation:

\begin{equation} \label{f1}
f^{(1)}(\mathbf{k}) = - \frac{2\pi\tau(k)}{v_F} \frac{e^2F^2}{\hbar^2}L_0\frac{\partial N_{ph}^{(0)}(q)}{\partial T} \left( \frac{\mathbf{k}}{k} \vec{\nabla} T\right) I_1(\tilde{k}),
\end{equation}
where $\tau(k)$ is the electron relaxation time and $I_1(\tilde{k})$ is the dimensionless integral over phonon wave vector:
\begin{multline} \label{I1}
I_1(\tilde{k})=\int_0^{1}d\tilde{q}d\theta \tilde{q} A_{\textbf{k},\textbf{q}}\exp^{-2z_0\tilde{q}\frac{\pi}{a_0}}\cos(\theta) \times\\
\left(f^{(0)}(\tilde{k}) - f^{(0)}(|\tilde{\mathbf{k}}+\tilde{\mathbf{q}}|) \right) \times \\
\left[ \delta(\tilde{k}+\tilde{\omega}_{ph}-|\tilde{\mathbf{k}}+\tilde{\mathbf{q}}|) - \delta(\tilde{k}-\tilde{\omega}_{ph}-|\tilde{\mathbf{k}}+\tilde{\mathbf{q}}|) \right],
\end{multline}
where $\tilde{\omega}_{ph} = \frac{\omega_{ph} a_0}{v_F \pi}$, $\tilde{\mathbf{k}} = \frac{\mathbf{k}a_0}{\pi}$ and $f^{(0)}(\tilde{k})=(1+\exp((\pi\hbar v_f \tilde{k}/a_0)-\varepsilon_F)/k_BT)^{-1}$. 
Fig. \ref{fig1a} shows a typical profile of the integral $I_1(\tilde{k})$ as a function of electron wave vector amplitude for various Fermi energies.

In the case of nearly elastic scattering of electrons by acoustic phonons the expression in the round brackets in \eqref{I1} yields $\pm\frac{\partial f^{(0)}(\varepsilon_k)}{\partial \varepsilon} \hbar \omega_{ph}(q) \approx \pm \hat{\delta}(\varepsilon_k-\varepsilon_F) \hbar \omega_{ph}(q)$, where $\hat{\delta}$ is a delta function with a broadening of the order of temperature.

In contrast with the case of acoustic phonons, here $f^{(1)}$ is not linear in $\frac{\partial f^{(0)}}{\partial \varepsilon}$. However, $f^{(1)}$ differs from zero for wave vectors $k$ close to elastic circle of the radius $k_F$ (see. fig. \ref{fig1a}). Therefore, in \eqref{f1} we can assume $\tau(k) = \tau(k_F) = const$ without significant loss of accuracy. Electron conductivity in graphene is directly related to electron transport relaxation time as \cite{RevModPhys.83.407,PhysRevB.81.195442}
\begin{equation} \label{sigma}
\sigma = \frac{e^2}{\hbar} \frac{v_F k_F \tau(k_F)}{\pi}.
\end{equation}

The SPP driven current is given by integral
\begin{equation} \label{jph}
j_{ph} = \frac{\pi^2 ev_F}{a_0^2} \int d\tilde{\mathbf{k}} f^{(1)}(\tilde{\mathbf{k}}).
\end{equation}
Substituting the electron distribution correction \eqref{f1} and dividing the obtained expression by conductivity \eqref{sigma} yields the SPP-drag thermopower:

\begin{equation} \label{SSPP}
S_\textrm{SPP} = \frac{\pi^4k_B}{e} \frac{F^2e^2}{\varepsilon_Fa_0k_BT} \frac{L_0}{a_0} \frac{\hbar \omega_{ph}}{k_BT}\frac{e^{\frac{\hbar \omega_{ph}}{k_BT}}}{\left( e^{\frac{\hbar \omega_{ph}}{k_BT}} - 1 \right)^2 } \cdot I_2,
\end{equation}
where
\begin{equation} \label{I2}
I_2=\frac{1}{2}\int_0^{1}\tilde{k}d\tilde{k}I_1(\tilde{k}).
\end{equation}
As usually the expression for the thermopower has a form of $\frac{k_B}{e}=86 \mu$VK$^{-1}$ times a dimensionless factor, which depends on $\varepsilon_F$, $T$ etc.

In Fig. \ref{fig1b} we have shown the dependence of $I_2$ on the Fermi energy in graphene sample. Due to the fact that $I_1(k) \approx \frac{\partial}{\partial \varepsilon}f^{(0)}(\varepsilon-\frac{1}{2}\hbar\omega_{ph})$, for actual values of the Fermi energy, temperature and phonon frequency $I_2$ was approximated via formula

\begin{equation} \label{I2_approx1}
I_{2} \simeq A_{0} \left( \frac{\varepsilon_F+\frac{1}{2}\hbar \omega_{ph}}{\varepsilon_0}  \right) \left( \frac{\varepsilon_F}{\varepsilon_0} \right)^{0.7} \left( \frac{\hbar \omega_{ph}}{\varepsilon_0} \right)^{0.7},
\end{equation}
where $A_{0} = 5 \cdot 10^{-6}$ and $\varepsilon_0 = 0.1$ eV, and sufficient accuracy was reached.

\section{Results and discussion}

Table 1 lists the values of the SPP drag thermopower for graphene on various substrates. The values $\varepsilon_F=0.5$\,eV and $T=300$\,K were used. Values of phonon energies and magnitudes of Fr\"olich coupling were adopted from \cite{PhysRevB.81.195442}.

\begin{table*}[!ht]
\centering
\caption{SPP drag thermopower in for graphene on SiO$_2$, HfO$_2$, SiC and h-BN. $\varepsilon_F=0.5$\,eV and T=300\,K. Values are given in $\mu$VK$^{-1}$.}
\begin{tabular}{ l c c c r }
\hline
\hline
  &SiO$_2$ & HfO$_2$ & SiC & h-BN\\
  \hline
 const $\tau_{ph}=10$\,ps (a=1) &1.7 & 1.1 & 1.1 & 0.6\\
 const $L_{ph}=1\,\mu$m (a=0) & $1.0\cdot 10^3$ & $3\cdot 10^3$ & $6\cdot 10^2$ & $3\cdot 10^2$\\
\hline
\hline
\end{tabular}
\end{table*}

For $\varepsilon_F=0.5$\,eV, $T=300$\,K and phonon lifetime $\tau_{ph}=10$\,ps one has $S_\textrm{SPP} \approx 2$\,$\mu $VK$^{-1}$ for graphene on SiO$_2$ substrate, which is one order of magnitude lower than the diffusion contribution. One sees that the $S_\textrm{SPP}$ is of the same order for other considered substrates.

\begin{figure*}
\centering     
\subfigure[]{\label{fig2a}\includegraphics[width=0.49\textwidth]{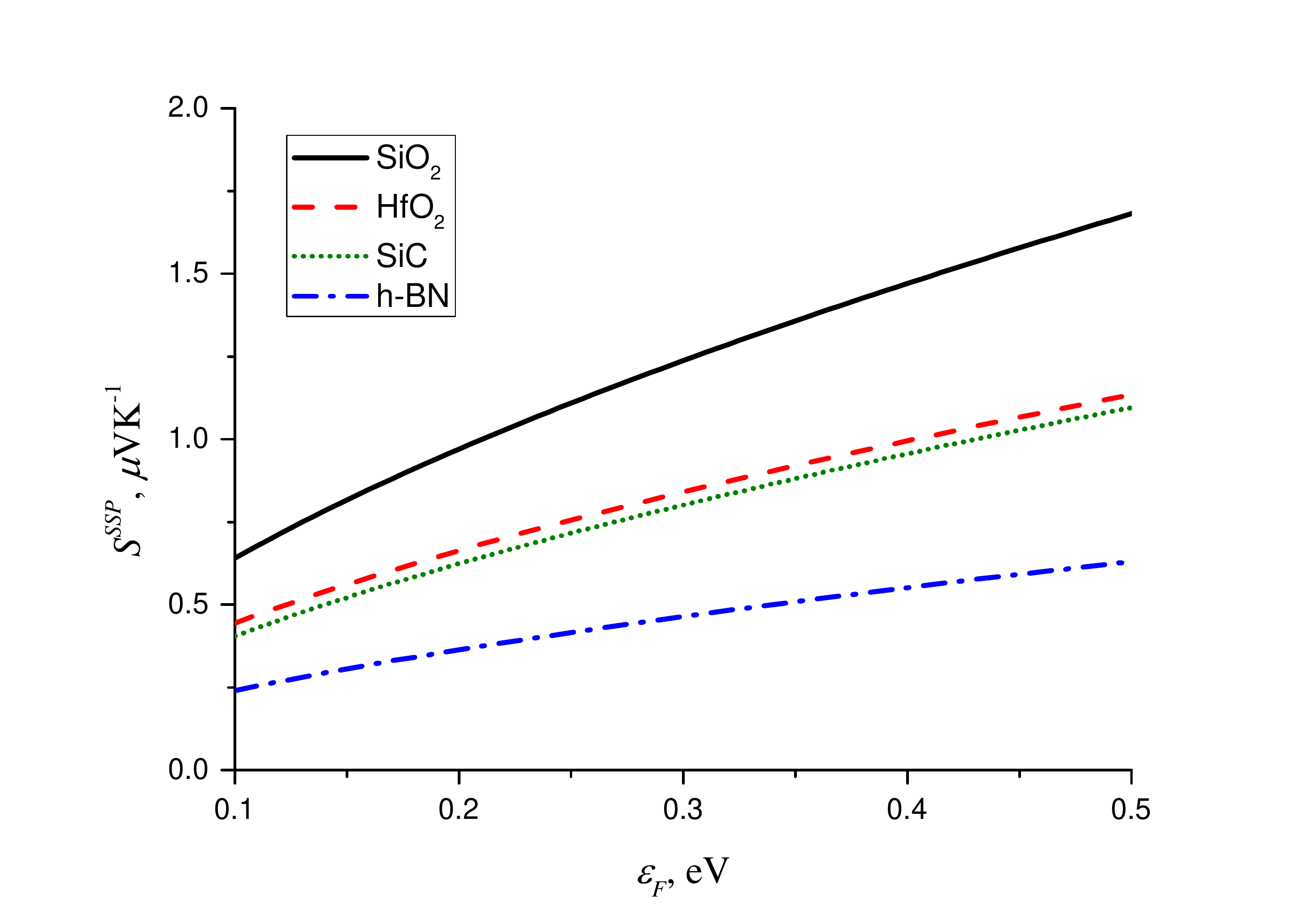}}
\subfigure[]{\label{fig2b}\includegraphics[width=0.49\textwidth]{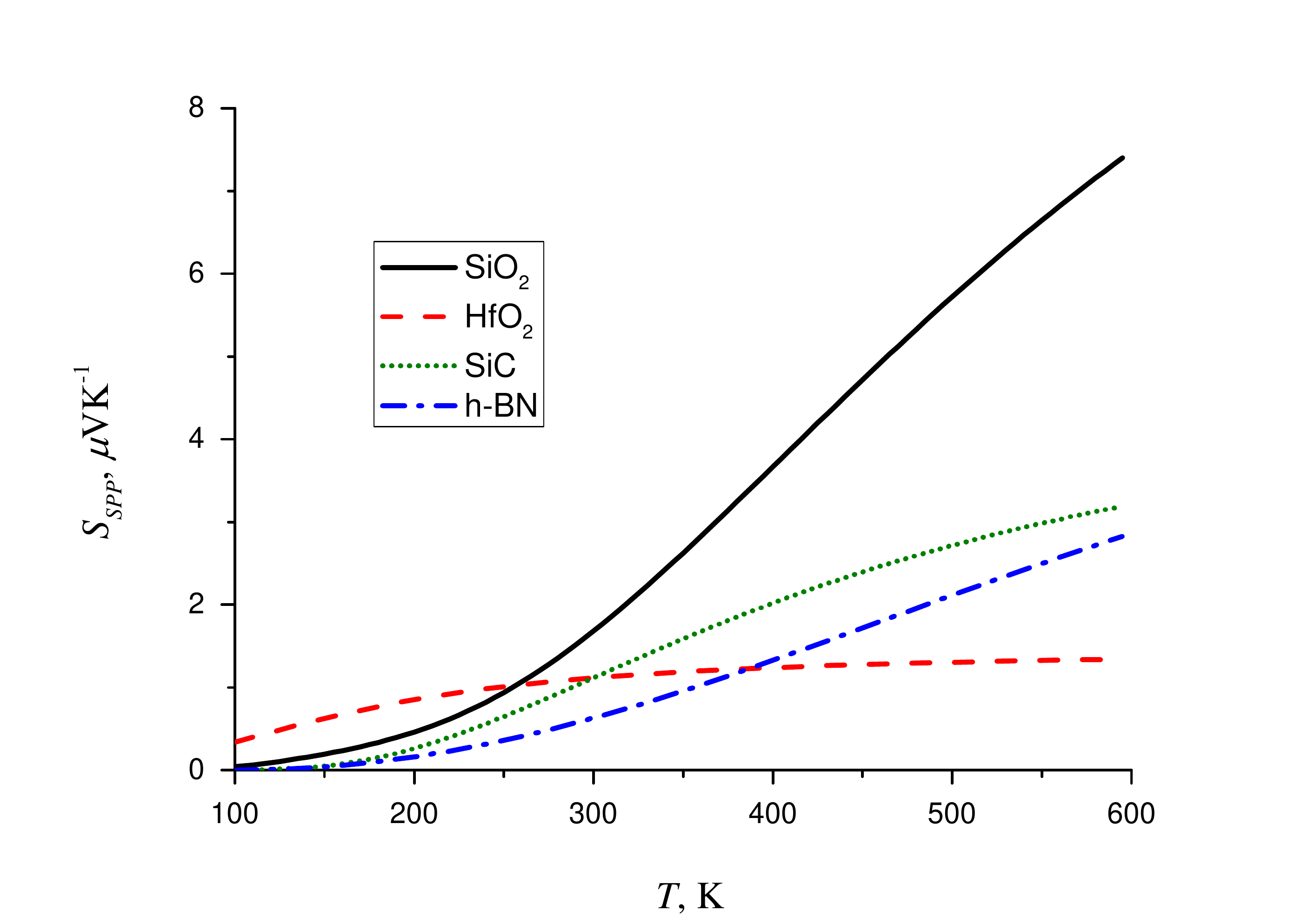}}
\caption{SPP drag thermopower  in graphene on various substrates. Constant phonon lifetime $\tau_{ph}$ is assumed to be 10\,ps. Panel (a). Dependence of $S_\textrm{SPP}$ on Fermi energy in graphene, $T=300$\,K. Panel (b). Dependence of $S_\textrm{SPP}$ on temperature for constant phonon lifetime, $\varepsilon_F=0.5$\,eV.}
\label{fig2}
\end{figure*}

Substitution of \eqref{I2_approx1} to \eqref{SSPP} allows us to conclude that the SPP thermopower moderately grows with Fermi energy in graphene. Due to exponential growth with temperature of the optical phonons occupation number, the SPP drag mechanism is only significant at relatively high temperatures. Fig. \ref{fig2} shows $S_{\textrm{SPP}}$ as a function of the Fermi energy and the temperature. By different dependencies on temperature and Fermi energy SPP contribution to the thermopower can be straightly distinguished from the ones of diffusion and intrinsic phonon drag mechanism.

\begin{figure}
\centering
\includegraphics[width=0.5\textwidth]{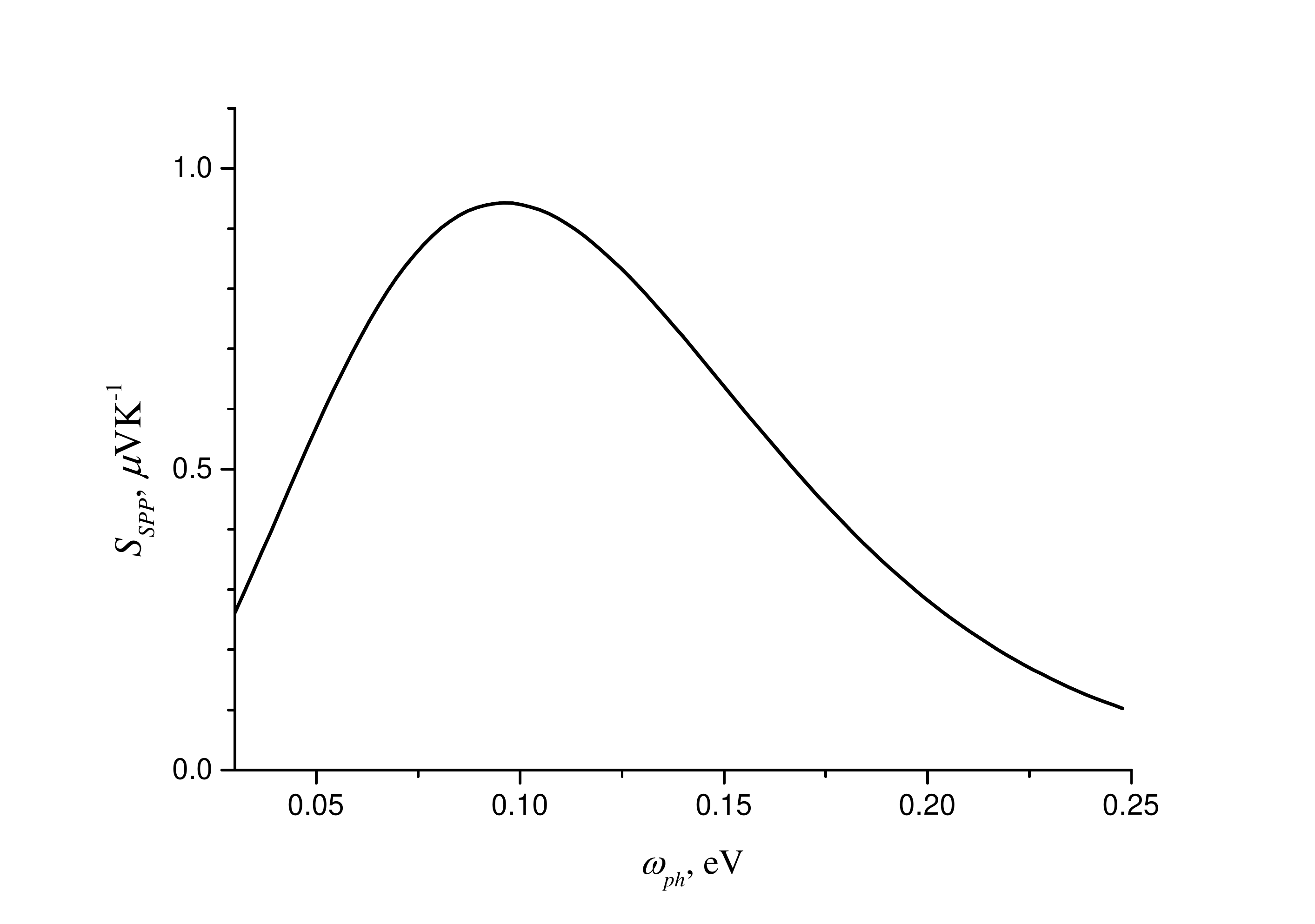}
\caption{SPP drag thermopower as a function of phonon energy. $T=300$\,K, $\varepsilon_F=0.5$\,eV. Phonon lifetime $\tau_{ph}=0.1$\,ps.}
\label{fig3}
\end{figure}

To suggest the substrate with maximal SPP drag effect it is important to analyze how $S_\textrm{\textrm{SPP}}$ depends on the optical phonon energy. Equation \eqref{Fsqared} shows that Fr\"olich coupling linearly grows with phonon energy and by averaging over dielectric constants of various materials one can write the following approximate relation: $F^2 = 0.005\hbar \omega_{ph}$. Substituting this to \eqref{SSPP} allows to plot $S_{\textrm{SPP}}$ as a function of energy of the substrate phonon. Fig. \ref{fig3} shows that the phonon energy most favorable for increasing of SPP drag thermopower is about 0.1\,eV. The energies of optical surface phonons in considered substrates are close to this value. It means that maximal achievable value of $S_\textrm{SPP}$ is reached for considered substrates.

It is interesting to compare coupling between electrons and intrinsic acoustic phonons with the one for SPPs. The coupling of electrons with intrinsic acoustic phonons in graphene reads as\cite{koniakhin2013phonon,bhargavi2013scattering}:
\begin{equation} \label{matrixElintr}
W_{\mathbf{k}\rightarrow \mathbf{k}+\mathbf{q}}^{intrinsic} = A_{\mathbf{k},\mathbf{q}} \frac{\hbar D^2q}{\rho Sv_s},
\end{equation}
where $v_s$ is the sound velocity in graphene.

For the phonon wave vector $q$ of the order of Fermi wave vector $k_F$ corresponding to $\varepsilon_F = 0.1$eV one can estimate $W^{SPP} / W^{intrinsic} \approx 600$ and for $\varepsilon_F = 0.5$\,eV one obtains $W^{SPP} / W^{intrinsic} \approx 15$. It means that electrostatic interaction of electrons with remote substrate phonons can be stronger than coupling with intrinsic phonons via deformation potential.

Nevertheless, the strong coupling between electrons in graphene and SPPs is not enough to overcome exponentially low occupation number of SPPs and short mean free path stemming from low group velocity of phonons with $q \approx k_F$. We conclude that the contribution to thermopower from drag of electrons by SPPs is weaker than the diffusion contribution, which explains current experimental data on the thermopower in graphene \cite{PhysRevB.80.081413,PhysRevB.83.113403,PhysRevLett.102.096807,PhysRevLett.102.166808}, being in agreement with Mott's law.

The effect of phonon drag will potentially play a role for a substrate with high optical phonon lifetime. Limitation of the phonon lifetime due to three-phonon processes will not allow dominance of SPP drag contribution in graphene thermopower with high probability. Moreover to obtain the predicted values of thermopower the interface between graphene sheet and substrate is to be thin and smooth. In case of mechanically exfoliated graphene the distance between substrate and graphene sheet can be expected to be higher than adopted here value. For epitaxial graphene dead buffer layer can also negatively effect on graphene properties \cite{qi2010epitaxial}.

The considered effect of SPP drag can be expected even for nonpolar substrates like diamond due to polarizability of interatomic bonds \cite{mahan2009kapitza}, which is important for creating composite nanocarbon-based thermoelectric devices. However we one sees that for nonpolar substrates the SPP drag thermopower will be smaller than for polar substrates considered in this paper.

\section{Acknowledgements}

This study was supported by Russian Science Foundation (grant \# 16-19-00075). The author is grateful to to E.D. Eidelman, who has encouraged present study. We are gratefully indebted to M.M. Glazov for fruitful discussions.

\bibliography{spp}

\begin{thebibliography}{58}%
\makeatletter
\providecommand \@ifxundefined [1]{%
 \@ifx{#1\undefined}
}%
\providecommand \@ifnum [1]{%
 \ifnum #1\expandafter \@firstoftwo
 \else \expandafter \@secondoftwo
 \fi
}%
\providecommand \@ifx [1]{%
 \ifx #1\expandafter \@firstoftwo
 \else \expandafter \@secondoftwo
 \fi
}%
\providecommand \natexlab [1]{#1}%
\providecommand \enquote  [1]{``#1''}%
\providecommand \bibnamefont  [1]{#1}%
\providecommand \bibfnamefont [1]{#1}%
\providecommand \citenamefont [1]{#1}%
\providecommand \href@noop [0]{\@secondoftwo}%
\providecommand \href [0]{\begingroup \@sanitize@url \@href}%
\providecommand \@href[1]{\@@startlink{#1}\@@href}%
\providecommand \@@href[1]{\endgroup#1\@@endlink}%
\providecommand \@sanitize@url [0]{\catcode `\\12\catcode `\$12\catcode
  `\&12\catcode `\#12\catcode `\^12\catcode `\_12\catcode `\%12\relax}%
\providecommand \@@startlink[1]{}%
\providecommand \@@endlink[0]{}%
\providecommand \url  [0]{\begingroup\@sanitize@url \@url }%
\providecommand \@url [1]{\endgroup\@href {#1}{\urlprefix }}%
\providecommand \urlprefix  [0]{URL }%
\providecommand \Eprint [0]{\href }%
\providecommand \doibase [0]{http://dx.doi.org/}%
\providecommand \selectlanguage [0]{\@gobble}%
\providecommand \bibinfo  [0]{\@secondoftwo}%
\providecommand \bibfield  [0]{\@secondoftwo}%
\providecommand \translation [1]{[#1]}%
\providecommand \BibitemOpen [0]{}%
\providecommand \bibitemStop [0]{}%
\providecommand \bibitemNoStop [0]{.\EOS\space}%
\providecommand \EOS [0]{\spacefactor3000\relax}%
\providecommand \BibitemShut  [1]{\csname bibitem#1\endcsname}%
\let\auto@bib@innerbib\@empty
\bibitem [{\citenamefont {Frank}\ \emph {et~al.}(2007)\citenamefont {Frank},
  \citenamefont {Tanenbaum}, \citenamefont {Van~der Zande},\ and\ \citenamefont
  {McEuen}}]{frank2007mechanical}%
  \BibitemOpen
  \bibfield  {author} {\bibinfo {author} {\bibfnamefont {I.}~\bibnamefont
  {Frank}}, \bibinfo {author} {\bibfnamefont {D.~M.}\ \bibnamefont
  {Tanenbaum}}, \bibinfo {author} {\bibfnamefont {A.}~\bibnamefont {Van~der
  Zande}}, \ and\ \bibinfo {author} {\bibfnamefont {P.~L.}\ \bibnamefont
  {McEuen}},\ }\href@noop {} {\bibfield  {journal} {\bibinfo  {journal}
  {Journal of Vacuum Science \& Technology B}\ }\textbf {\bibinfo {volume}
  {25}},\ \bibinfo {pages} {2558} (\bibinfo {year} {2007})}\BibitemShut
  {NoStop}%
\bibitem [{\citenamefont {Lee}\ \emph {et~al.}(2008)\citenamefont {Lee},
  \citenamefont {Wei}, \citenamefont {Kysar},\ and\ \citenamefont
  {Hone}}]{lee2008measurement}%
  \BibitemOpen
  \bibfield  {author} {\bibinfo {author} {\bibfnamefont {C.}~\bibnamefont
  {Lee}}, \bibinfo {author} {\bibfnamefont {X.}~\bibnamefont {Wei}}, \bibinfo
  {author} {\bibfnamefont {J.~W.}\ \bibnamefont {Kysar}}, \ and\ \bibinfo
  {author} {\bibfnamefont {J.}~\bibnamefont {Hone}},\ }\href@noop {} {\bibfield
   {journal} {\bibinfo  {journal} {science}\ }\textbf {\bibinfo {volume}
  {321}},\ \bibinfo {pages} {385} (\bibinfo {year} {2008})}\BibitemShut
  {NoStop}%
\bibitem [{\citenamefont {Castro~Neto}\ \emph {et~al.}(2009)\citenamefont
  {Castro~Neto}, \citenamefont {Guinea}, \citenamefont {Peres}, \citenamefont
  {Novoselov},\ and\ \citenamefont {Geim}}]{RevModPhys.81.109}%
  \BibitemOpen
  \bibfield  {author} {\bibinfo {author} {\bibfnamefont {A.~H.}\ \bibnamefont
  {Castro~Neto}}, \bibinfo {author} {\bibfnamefont {F.}~\bibnamefont {Guinea}},
  \bibinfo {author} {\bibfnamefont {N.~M.~R.}\ \bibnamefont {Peres}}, \bibinfo
  {author} {\bibfnamefont {K.~S.}\ \bibnamefont {Novoselov}}, \ and\ \bibinfo
  {author} {\bibfnamefont {A.~K.}\ \bibnamefont {Geim}},\ }\href {\doibase
  10.1103/RevModPhys.81.109} {\bibfield  {journal} {\bibinfo  {journal} {Rev.
  Mod. Phys.}\ }\textbf {\bibinfo {volume} {81}},\ \bibinfo {pages} {109}
  (\bibinfo {year} {2009})}\BibitemShut {NoStop}%
\bibitem [{\citenamefont {Das~Sarma}\ \emph {et~al.}(2011)\citenamefont
  {Das~Sarma}, \citenamefont {Adam}, \citenamefont {Hwang},\ and\ \citenamefont
  {Rossi}}]{RevModPhys.83.407}%
  \BibitemOpen
  \bibfield  {author} {\bibinfo {author} {\bibfnamefont {S.}~\bibnamefont
  {Das~Sarma}}, \bibinfo {author} {\bibfnamefont {S.}~\bibnamefont {Adam}},
  \bibinfo {author} {\bibfnamefont {E.~H.}\ \bibnamefont {Hwang}}, \ and\
  \bibinfo {author} {\bibfnamefont {E.}~\bibnamefont {Rossi}},\ }\href
  {\doibase 10.1103/RevModPhys.83.407} {\bibfield  {journal} {\bibinfo
  {journal} {Rev. Mod. Phys.}\ }\textbf {\bibinfo {volume} {83}},\ \bibinfo
  {pages} {407} (\bibinfo {year} {2011})}\BibitemShut {NoStop}%
\bibitem [{\citenamefont {Hwang}\ \emph {et~al.}(2007)\citenamefont {Hwang},
  \citenamefont {Adam},\ and\ \citenamefont
  {Das~Sarma}}]{PhysRevLett.98.186806}%
  \BibitemOpen
  \bibfield  {author} {\bibinfo {author} {\bibfnamefont {E.~H.}\ \bibnamefont
  {Hwang}}, \bibinfo {author} {\bibfnamefont {S.}~\bibnamefont {Adam}}, \ and\
  \bibinfo {author} {\bibfnamefont {S.}~\bibnamefont {Das~Sarma}},\ }\href
  {\doibase 10.1103/PhysRevLett.98.186806} {\bibfield  {journal} {\bibinfo
  {journal} {Phys. Rev. Lett.}\ }\textbf {\bibinfo {volume} {98}},\ \bibinfo
  {pages} {186806} (\bibinfo {year} {2007})}\BibitemShut {NoStop}%
\bibitem [{\citenamefont {Nalitov}\ \emph {et~al.}(2012)\citenamefont
  {Nalitov}, \citenamefont {Golub},\ and\ \citenamefont
  {Ivchenko}}]{PhysRevB.86.115301}%
  \BibitemOpen
  \bibfield  {author} {\bibinfo {author} {\bibfnamefont {A.~V.}\ \bibnamefont
  {Nalitov}}, \bibinfo {author} {\bibfnamefont {L.~E.}\ \bibnamefont {Golub}},
  \ and\ \bibinfo {author} {\bibfnamefont {E.~L.}\ \bibnamefont {Ivchenko}},\
  }\href {\doibase 10.1103/PhysRevB.86.115301} {\bibfield  {journal} {\bibinfo
  {journal} {Phys. Rev. B}\ }\textbf {\bibinfo {volume} {86}},\ \bibinfo
  {pages} {115301} (\bibinfo {year} {2012})}\BibitemShut {NoStop}%
\bibitem [{\citenamefont {Glazov}\ and\ \citenamefont
  {Ganichev}(2014)}]{Glazov2014101}%
  \BibitemOpen
  \bibfield  {author} {\bibinfo {author} {\bibfnamefont {M.}~\bibnamefont
  {Glazov}}\ and\ \bibinfo {author} {\bibfnamefont {S.}~\bibnamefont
  {Ganichev}},\ }\href {\doibase
  http://dx.doi.org/10.1016/j.physrep.2013.10.003} {\bibfield  {journal}
  {\bibinfo  {journal} {Physics Reports}\ }\textbf {\bibinfo {volume} {535}},\
  \bibinfo {pages} {101 } (\bibinfo {year} {2014})},\ \bibinfo {note} {high
  frequency electric field induced nonlinear effects in graphene}\BibitemShut
  {NoStop}%
\bibitem [{\citenamefont {Karch}\ \emph {et~al.}(2011)\citenamefont {Karch},
  \citenamefont {Drexler}, \citenamefont {Olbrich}, \citenamefont
  {Fehrenbacher}, \citenamefont {Hirmer}, \citenamefont {Glazov}, \citenamefont
  {Tarasenko}, \citenamefont {Ivchenko}, \citenamefont {Birkner}, \citenamefont
  {Eroms}, \citenamefont {Weiss}, \citenamefont {Yakimova}, \citenamefont
  {Lara-Avila}, \citenamefont {Kubatkin}, \citenamefont {Ostler}, \citenamefont
  {Seyller},\ and\ \citenamefont {Ganichev}}]{PhysRevLett.107.276601}%
  \BibitemOpen
  \bibfield  {author} {\bibinfo {author} {\bibfnamefont {J.}~\bibnamefont
  {Karch}}, \bibinfo {author} {\bibfnamefont {C.}~\bibnamefont {Drexler}},
  \bibinfo {author} {\bibfnamefont {P.}~\bibnamefont {Olbrich}}, \bibinfo
  {author} {\bibfnamefont {M.}~\bibnamefont {Fehrenbacher}}, \bibinfo {author}
  {\bibfnamefont {M.}~\bibnamefont {Hirmer}}, \bibinfo {author} {\bibfnamefont
  {M.~M.}\ \bibnamefont {Glazov}}, \bibinfo {author} {\bibfnamefont {S.~A.}\
  \bibnamefont {Tarasenko}}, \bibinfo {author} {\bibfnamefont {E.~L.}\
  \bibnamefont {Ivchenko}}, \bibinfo {author} {\bibfnamefont {B.}~\bibnamefont
  {Birkner}}, \bibinfo {author} {\bibfnamefont {J.}~\bibnamefont {Eroms}},
  \bibinfo {author} {\bibfnamefont {D.}~\bibnamefont {Weiss}}, \bibinfo
  {author} {\bibfnamefont {R.}~\bibnamefont {Yakimova}}, \bibinfo {author}
  {\bibfnamefont {S.}~\bibnamefont {Lara-Avila}}, \bibinfo {author}
  {\bibfnamefont {S.}~\bibnamefont {Kubatkin}}, \bibinfo {author}
  {\bibfnamefont {M.}~\bibnamefont {Ostler}}, \bibinfo {author} {\bibfnamefont
  {T.}~\bibnamefont {Seyller}}, \ and\ \bibinfo {author} {\bibfnamefont
  {S.~D.}\ \bibnamefont {Ganichev}},\ }\href {\doibase
  10.1103/PhysRevLett.107.276601} {\bibfield  {journal} {\bibinfo  {journal}
  {Phys. Rev. Lett.}\ }\textbf {\bibinfo {volume} {107}},\ \bibinfo {pages}
  {276601} (\bibinfo {year} {2011})}\BibitemShut {NoStop}%
\bibitem [{\citenamefont {Alofi}\ and\ \citenamefont
  {Srivastava}(2012)}]{6235226}%
  \BibitemOpen
  \bibfield  {author} {\bibinfo {author} {\bibfnamefont {A.}~\bibnamefont
  {Alofi}}\ and\ \bibinfo {author} {\bibfnamefont {G.~P.}\ \bibnamefont
  {Srivastava}},\ }\href {\doibase 10.1063/1.4733690} {\bibfield  {journal}
  {\bibinfo  {journal} {Journal of Applied Physics}\ }\textbf {\bibinfo
  {volume} {112}},\ \bibinfo {pages} {013517 } (\bibinfo {year}
  {2012})}\BibitemShut {NoStop}%
\bibitem [{\citenamefont {Nika}\ \emph {et~al.}(2009)\citenamefont {Nika},
  \citenamefont {Pokatilov}, \citenamefont {Askerov},\ and\ \citenamefont
  {Balandin}}]{nika2009phonon}%
  \BibitemOpen
  \bibfield  {author} {\bibinfo {author} {\bibfnamefont {D.}~\bibnamefont
  {Nika}}, \bibinfo {author} {\bibfnamefont {E.}~\bibnamefont {Pokatilov}},
  \bibinfo {author} {\bibfnamefont {A.}~\bibnamefont {Askerov}}, \ and\
  \bibinfo {author} {\bibfnamefont {A.}~\bibnamefont {Balandin}},\ }\href@noop
  {} {\bibfield  {journal} {\bibinfo  {journal} {Physical Review B}\ }\textbf
  {\bibinfo {volume} {79}},\ \bibinfo {pages} {155413} (\bibinfo {year}
  {2009})}\BibitemShut {NoStop}%
\bibitem [{\citenamefont {Nika}\ and\ \citenamefont
  {Balandin}(2012)}]{nika2012two}%
  \BibitemOpen
  \bibfield  {author} {\bibinfo {author} {\bibfnamefont {D.~L.}\ \bibnamefont
  {Nika}}\ and\ \bibinfo {author} {\bibfnamefont {A.~A.}\ \bibnamefont
  {Balandin}},\ }\href@noop {} {\bibfield  {journal} {\bibinfo  {journal}
  {Journal of Physics: Condensed Matter}\ }\textbf {\bibinfo {volume} {24}},\
  \bibinfo {pages} {233203} (\bibinfo {year} {2012})}\BibitemShut {NoStop}%
\bibitem [{\citenamefont {Checkelsky}\ and\ \citenamefont
  {Ong}(2009)}]{PhysRevB.80.081413}%
  \BibitemOpen
  \bibfield  {author} {\bibinfo {author} {\bibfnamefont {J.~G.}\ \bibnamefont
  {Checkelsky}}\ and\ \bibinfo {author} {\bibfnamefont {N.~P.}\ \bibnamefont
  {Ong}},\ }\href {\doibase 10.1103/PhysRevB.80.081413} {\bibfield  {journal}
  {\bibinfo  {journal} {Phys. Rev. B}\ }\textbf {\bibinfo {volume} {80}},\
  \bibinfo {pages} {081413} (\bibinfo {year} {2009})}\BibitemShut {NoStop}%
\bibitem [{\citenamefont {Wang}\ and\ \citenamefont
  {Shi}(2011)}]{PhysRevB.83.113403}%
  \BibitemOpen
  \bibfield  {author} {\bibinfo {author} {\bibfnamefont {D.}~\bibnamefont
  {Wang}}\ and\ \bibinfo {author} {\bibfnamefont {J.}~\bibnamefont {Shi}},\
  }\href {\doibase 10.1103/PhysRevB.83.113403} {\bibfield  {journal} {\bibinfo
  {journal} {Phys. Rev. B}\ }\textbf {\bibinfo {volume} {83}},\ \bibinfo
  {pages} {113403} (\bibinfo {year} {2011})}\BibitemShut {NoStop}%
\bibitem [{\citenamefont {Zuev}\ \emph {et~al.}(2009)\citenamefont {Zuev},
  \citenamefont {Chang},\ and\ \citenamefont {Kim}}]{PhysRevLett.102.096807}%
  \BibitemOpen
  \bibfield  {author} {\bibinfo {author} {\bibfnamefont {Y.~M.}\ \bibnamefont
  {Zuev}}, \bibinfo {author} {\bibfnamefont {W.}~\bibnamefont {Chang}}, \ and\
  \bibinfo {author} {\bibfnamefont {P.}~\bibnamefont {Kim}},\ }\href {\doibase
  10.1103/PhysRevLett.102.096807} {\bibfield  {journal} {\bibinfo  {journal}
  {Phys. Rev. Lett.}\ }\textbf {\bibinfo {volume} {102}},\ \bibinfo {pages}
  {096807} (\bibinfo {year} {2009})}\BibitemShut {NoStop}%
\bibitem [{\citenamefont {Wei}\ \emph {et~al.}(2009)\citenamefont {Wei},
  \citenamefont {Bao}, \citenamefont {Pu}, \citenamefont {Lau},\ and\
  \citenamefont {Shi}}]{PhysRevLett.102.166808}%
  \BibitemOpen
  \bibfield  {author} {\bibinfo {author} {\bibfnamefont {P.}~\bibnamefont
  {Wei}}, \bibinfo {author} {\bibfnamefont {W.}~\bibnamefont {Bao}}, \bibinfo
  {author} {\bibfnamefont {Y.}~\bibnamefont {Pu}}, \bibinfo {author}
  {\bibfnamefont {C.~N.}\ \bibnamefont {Lau}}, \ and\ \bibinfo {author}
  {\bibfnamefont {J.}~\bibnamefont {Shi}},\ }\href {\doibase
  10.1103/PhysRevLett.102.166808} {\bibfield  {journal} {\bibinfo  {journal}
  {Phys. Rev. Lett.}\ }\textbf {\bibinfo {volume} {102}},\ \bibinfo {pages}
  {166808} (\bibinfo {year} {2009})}\BibitemShut {NoStop}%
\bibitem [{\citenamefont {Dollfus}\ \emph {et~al.}(2015)\citenamefont
  {Dollfus}, \citenamefont {Nguyen} \emph
  {et~al.}}]{dollfus2015thermoelectric}%
  \BibitemOpen
  \bibfield  {author} {\bibinfo {author} {\bibfnamefont {P.}~\bibnamefont
  {Dollfus}}, \bibinfo {author} {\bibfnamefont {V.~H.}\ \bibnamefont {Nguyen}},
   \emph {et~al.},\ }\href@noop {} {\bibfield  {journal} {\bibinfo  {journal}
  {Journal of Physics: Condensed Matter}\ }\textbf {\bibinfo {volume} {27}},\
  \bibinfo {pages} {133204} (\bibinfo {year} {2015})}\BibitemShut {NoStop}%
\bibitem [{\citenamefont {Kubakaddi}(2009)}]{PhysRevB.79.075417}%
  \BibitemOpen
  \bibfield  {author} {\bibinfo {author} {\bibfnamefont {S.~S.}\ \bibnamefont
  {Kubakaddi}},\ }\href {\doibase 10.1103/PhysRevB.79.075417} {\bibfield
  {journal} {\bibinfo  {journal} {Phys. Rev. B}\ }\textbf {\bibinfo {volume}
  {79}},\ \bibinfo {pages} {075417} (\bibinfo {year} {2009})}\BibitemShut
  {NoStop}%
\bibitem [{\citenamefont {Hwang}\ \emph {et~al.}(2009)\citenamefont {Hwang},
  \citenamefont {Rossi},\ and\ \citenamefont {Das~Sarma}}]{PhysRevB.80.235415}%
  \BibitemOpen
  \bibfield  {author} {\bibinfo {author} {\bibfnamefont {E.~H.}\ \bibnamefont
  {Hwang}}, \bibinfo {author} {\bibfnamefont {E.}~\bibnamefont {Rossi}}, \ and\
  \bibinfo {author} {\bibfnamefont {S.}~\bibnamefont {Das~Sarma}},\ }\href
  {\doibase 10.1103/PhysRevB.80.235415} {\bibfield  {journal} {\bibinfo
  {journal} {Phys. Rev. B}\ }\textbf {\bibinfo {volume} {80}},\ \bibinfo
  {pages} {235415} (\bibinfo {year} {2009})}\BibitemShut {NoStop}%
\bibitem [{\citenamefont {Kubakaddi}\ and\ \citenamefont
  {Bhargavi}(2010)}]{PhysRevB.82.155410}%
  \BibitemOpen
  \bibfield  {author} {\bibinfo {author} {\bibfnamefont {S.~S.}\ \bibnamefont
  {Kubakaddi}}\ and\ \bibinfo {author} {\bibfnamefont {K.~S.}\ \bibnamefont
  {Bhargavi}},\ }\href {\doibase 10.1103/PhysRevB.82.155410} {\bibfield
  {journal} {\bibinfo  {journal} {Phys. Rev. B}\ }\textbf {\bibinfo {volume}
  {82}},\ \bibinfo {pages} {155410} (\bibinfo {year} {2010})}\BibitemShut
  {NoStop}%
\bibitem [{\citenamefont {Vaidya}\ \emph {et~al.}(2012)\citenamefont {Vaidya},
  \citenamefont {Sankeshwar},\ and\ \citenamefont {Mulimani}}]{vaidya:093711}%
  \BibitemOpen
  \bibfield  {author} {\bibinfo {author} {\bibfnamefont {R.~G.}\ \bibnamefont
  {Vaidya}}, \bibinfo {author} {\bibfnamefont {N.~S.}\ \bibnamefont
  {Sankeshwar}}, \ and\ \bibinfo {author} {\bibfnamefont {B.~G.}\ \bibnamefont
  {Mulimani}},\ }\href {\doibase 10.1063/1.4764335} {\bibfield  {journal}
  {\bibinfo  {journal} {Journal of Applied Physics}\ }\textbf {\bibinfo
  {volume} {112}},\ \bibinfo {eid} {093711} (\bibinfo {year}
  {2012})}\BibitemShut {NoStop}%
\bibitem [{\citenamefont {Sankeshwar}\ \emph {et~al.}(2013)\citenamefont
  {Sankeshwar}, \citenamefont {Vaidya},\ and\ \citenamefont
  {Mulimani}}]{PSSB:PSSB201248302}%
  \BibitemOpen
  \bibfield  {author} {\bibinfo {author} {\bibfnamefont {N.~S.}\ \bibnamefont
  {Sankeshwar}}, \bibinfo {author} {\bibfnamefont {R.~G.}\ \bibnamefont
  {Vaidya}}, \ and\ \bibinfo {author} {\bibfnamefont {B.~G.}\ \bibnamefont
  {Mulimani}},\ }\href {\doibase 10.1002/pssb.201248302} {\bibfield  {journal}
  {\bibinfo  {journal} {physica status solidi (b)}\ }\textbf {\bibinfo {volume}
  {250}},\ \bibinfo {pages} {1356} (\bibinfo {year} {2013})}\BibitemShut
  {NoStop}%
\bibitem [{\citenamefont {Koniakhin}\ and\ \citenamefont
  {Eidelman}(2013)}]{koniakhin2013phonon}%
  \BibitemOpen
  \bibfield  {author} {\bibinfo {author} {\bibfnamefont {S.}~\bibnamefont
  {Koniakhin}}\ and\ \bibinfo {author} {\bibfnamefont {E.}~\bibnamefont
  {Eidelman}},\ }\href@noop {} {\bibfield  {journal} {\bibinfo  {journal} {EPL
  (Europhysics Letters)}\ }\textbf {\bibinfo {volume} {103}},\ \bibinfo {pages}
  {37006} (\bibinfo {year} {2013})}\BibitemShut {NoStop}%
\bibitem [{\citenamefont {Alisultanov}(2015)}]{alisultanov2015thermodynamics}%
  \BibitemOpen
  \bibfield  {author} {\bibinfo {author} {\bibfnamefont {Z.}~\bibnamefont
  {Alisultanov}},\ }\href@noop {} {\bibfield  {journal} {\bibinfo  {journal}
  {Physica E: Low-dimensional Systems and Nanostructures}\ }\textbf {\bibinfo
  {volume} {69}},\ \bibinfo {pages} {89} (\bibinfo {year} {2015})}\BibitemShut
  {NoStop}%
\bibitem [{\citenamefont {Alisultanov}\ and\ \citenamefont
  {Mirzegasanova}(2014)}]{alisultanov2014anomalous}%
  \BibitemOpen
  \bibfield  {author} {\bibinfo {author} {\bibfnamefont {Z.}~\bibnamefont
  {Alisultanov}}\ and\ \bibinfo {author} {\bibfnamefont {N.}~\bibnamefont
  {Mirzegasanova}},\ }\href@noop {} {\bibfield  {journal} {\bibinfo  {journal}
  {Technical Physics}\ }\textbf {\bibinfo {volume} {59}},\ \bibinfo {pages}
  {1562} (\bibinfo {year} {2014})}\BibitemShut {NoStop}%
\bibitem [{\citenamefont {Gurevich}(1946)}]{Gurevich46}%
  \BibitemOpen
  \bibfield  {author} {\bibinfo {author} {\bibfnamefont {L.}~\bibnamefont
  {Gurevich}},\ }\href@noop {} {\bibfield  {journal} {\bibinfo  {journal} {J.
  Phys. (USSR)}\ }\textbf {\bibinfo {volume} {9}},\ \bibinfo {pages} {4}
  (\bibinfo {year} {1946})}\BibitemShut {NoStop}%
\bibitem [{\citenamefont {Eidelman}\ and\ \citenamefont {Vul}(2007)}]{EidVul}%
  \BibitemOpen
  \bibfield  {author} {\bibinfo {author} {\bibfnamefont {E.~D.}\ \bibnamefont
  {Eidelman}}\ and\ \bibinfo {author} {\bibfnamefont {A.~Y.}\ \bibnamefont
  {Vul}},\ }\href {http://stacks.iop.org/0953-8984/19/i=26/a=266210} {\bibfield
   {journal} {\bibinfo  {journal} {Journal of Physics: Condensed Matter}\
  }\textbf {\bibinfo {volume} {19}},\ \bibinfo {pages} {266210} (\bibinfo
  {year} {2007})}\BibitemShut {NoStop}%
\bibitem [{\citenamefont {Ayache}\ \emph {et~al.}(1980)\citenamefont {Ayache},
  \citenamefont {de~Combarieu},\ and\ \citenamefont
  {Jay-Gerin}}]{PhysRevB.21.2462}%
  \BibitemOpen
  \bibfield  {author} {\bibinfo {author} {\bibfnamefont {C.}~\bibnamefont
  {Ayache}}, \bibinfo {author} {\bibfnamefont {A.}~\bibnamefont
  {de~Combarieu}}, \ and\ \bibinfo {author} {\bibfnamefont {J.~P.}\
  \bibnamefont {Jay-Gerin}},\ }\href {\doibase 10.1103/PhysRevB.21.2462}
  {\bibfield  {journal} {\bibinfo  {journal} {Phys. Rev. B}\ }\textbf {\bibinfo
  {volume} {21}},\ \bibinfo {pages} {2462} (\bibinfo {year}
  {1980})}\BibitemShut {NoStop}%
\bibitem [{\citenamefont {Sugihara}(1983)}]{PhysRevB.28.2157}%
  \BibitemOpen
  \bibfield  {author} {\bibinfo {author} {\bibfnamefont {K.}~\bibnamefont
  {Sugihara}},\ }\href {\doibase 10.1103/PhysRevB.28.2157} {\bibfield
  {journal} {\bibinfo  {journal} {Phys. Rev. B}\ }\textbf {\bibinfo {volume}
  {28}},\ \bibinfo {pages} {2157} (\bibinfo {year} {1983})}\BibitemShut
  {NoStop}%
\bibitem [{\citenamefont {Sugihara}\ \emph {et~al.}(1986)\citenamefont
  {Sugihara}, \citenamefont {Hishiyama},\ and\ \citenamefont
  {Ono}}]{PhysRevB.34.4298}%
  \BibitemOpen
  \bibfield  {author} {\bibinfo {author} {\bibfnamefont {K.}~\bibnamefont
  {Sugihara}}, \bibinfo {author} {\bibfnamefont {Y.}~\bibnamefont {Hishiyama}},
  \ and\ \bibinfo {author} {\bibfnamefont {A.}~\bibnamefont {Ono}},\ }\href
  {\doibase 10.1103/PhysRevB.34.4298} {\bibfield  {journal} {\bibinfo
  {journal} {Phys. Rev. B}\ }\textbf {\bibinfo {volume} {34}},\ \bibinfo
  {pages} {4298} (\bibinfo {year} {1986})}\BibitemShut {NoStop}%
\bibitem [{\citenamefont {R.}\ \emph {et~al.}(2010)\citenamefont {R.},
  \citenamefont {F.}, \citenamefont {MericI.}, \citenamefont {LeeC.},
  \citenamefont {WangL.}, \citenamefont {SorgenfreiS.}, \citenamefont
  {WatanabeK.}, \citenamefont {TaniguchiT.}, \citenamefont {KimP.},
  \citenamefont {L.},\ and\ \citenamefont {HoneJ.}}]{R.2010}%
  \BibitemOpen
  \bibfield  {author} {\bibinfo {author} {\bibfnamefont {D.}~\bibnamefont
  {R.}}, \bibinfo {author} {\bibfnamefont {Y.}~\bibnamefont {F.}}, \bibinfo
  {author} {\bibnamefont {MericI.}}, \bibinfo {author} {\bibnamefont {LeeC.}},
  \bibinfo {author} {\bibnamefont {WangL.}}, \bibinfo {author} {\bibnamefont
  {SorgenfreiS.}}, \bibinfo {author} {\bibnamefont {WatanabeK.}}, \bibinfo
  {author} {\bibnamefont {TaniguchiT.}}, \bibinfo {author} {\bibnamefont
  {KimP.}}, \bibinfo {author} {\bibfnamefont {S.}~\bibnamefont {L.}}, \ and\
  \bibinfo {author} {\bibnamefont {HoneJ.}},\ }\href {\doibase
  10.1038/nnano.2010.172} {\bibfield  {journal} {\bibinfo  {journal} {Nat
  Nano}\ }\textbf {\bibinfo {volume} {5}},\ \bibinfo {pages} {722} (\bibinfo
  {year} {2010})}\BibitemShut {NoStop}%
\bibitem [{\citenamefont {Frank}\ \emph {et~al.}(2014)\citenamefont {Frank},
  \citenamefont {Vejpravova}, \citenamefont {Holy}, \citenamefont {Kavan},\
  and\ \citenamefont {Kalbac}}]{Frank2014440}%
  \BibitemOpen
  \bibfield  {author} {\bibinfo {author} {\bibfnamefont {O.}~\bibnamefont
  {Frank}}, \bibinfo {author} {\bibfnamefont {J.}~\bibnamefont {Vejpravova}},
  \bibinfo {author} {\bibfnamefont {V.}~\bibnamefont {Holy}}, \bibinfo {author}
  {\bibfnamefont {L.}~\bibnamefont {Kavan}}, \ and\ \bibinfo {author}
  {\bibfnamefont {M.}~\bibnamefont {Kalbac}},\ }\href {\doibase
  http://dx.doi.org/10.1016/j.carbon.2013.11.020} {\bibfield  {journal}
  {\bibinfo  {journal} {Carbon}\ }\textbf {\bibinfo {volume} {68}},\ \bibinfo
  {pages} {440 } (\bibinfo {year} {2014})}\BibitemShut {NoStop}%
\bibitem [{\citenamefont {Wang}\ \emph {et~al.}(2008)\citenamefont {Wang},
  \citenamefont {Ni}, \citenamefont {Yu}, \citenamefont {Shen}, \citenamefont
  {Wang}, \citenamefont {Wu}, \citenamefont {Chen},\ and\ \citenamefont
  {Shen~Wee}}]{doi:10.1021/jp8008404}%
  \BibitemOpen
  \bibfield  {author} {\bibinfo {author} {\bibfnamefont {Y.~y.}\ \bibnamefont
  {Wang}}, \bibinfo {author} {\bibfnamefont {Z.~h.}\ \bibnamefont {Ni}},
  \bibinfo {author} {\bibfnamefont {T.}~\bibnamefont {Yu}}, \bibinfo {author}
  {\bibfnamefont {Z.~X.}\ \bibnamefont {Shen}}, \bibinfo {author}
  {\bibfnamefont {H.~m.}\ \bibnamefont {Wang}}, \bibinfo {author}
  {\bibfnamefont {Y.~h.}\ \bibnamefont {Wu}}, \bibinfo {author} {\bibfnamefont
  {W.}~\bibnamefont {Chen}}, \ and\ \bibinfo {author} {\bibfnamefont {A.~T.}\
  \bibnamefont {Shen~Wee}},\ }\href {\doibase 10.1021/jp8008404} {\bibfield
  {journal} {\bibinfo  {journal} {The Journal of Physical Chemistry C}\
  }\textbf {\bibinfo {volume} {112}},\ \bibinfo {pages} {10637} (\bibinfo
  {year} {2008})},\ \Eprint
  {http://arxiv.org/abs/http://pubs.acs.org/doi/pdf/10.1021/jp8008404}
  {http://pubs.acs.org/doi/pdf/10.1021/jp8008404} \BibitemShut {NoStop}%
\bibitem [{\citenamefont {Brako}\ \emph {et~al.}(2010)\citenamefont {Brako},
  \citenamefont {Å $\check{S}$ok$\check{c}$evi$\acute{c}$}, \citenamefont
  {Lazi$\acute{c}$},\ and\ \citenamefont {Atodiresei}}]{GrapheneIr111}%
  \BibitemOpen
  \bibfield  {author} {\bibinfo {author} {\bibfnamefont {R.}~\bibnamefont
  {Brako}}, \bibinfo {author} {\bibfnamefont {D.}~\bibnamefont
  {Å $\check{S}$ok$\check{c}$evi$\acute{c}$}}, \bibinfo {author} {\bibfnamefont
  {P.}~\bibnamefont {Lazi$\acute{c}$}}, \ and\ \bibinfo {author} {\bibfnamefont
  {N.}~\bibnamefont {Atodiresei}},\ }\href
  {http://stacks.iop.org/1367-2630/12/i=11/a=113016} {\bibfield  {journal}
  {\bibinfo  {journal} {New Journal of Physics}\ }\textbf {\bibinfo {volume}
  {12}},\ \bibinfo {pages} {113016} (\bibinfo {year} {2010})}\BibitemShut
  {NoStop}%
\bibitem [{\citenamefont {Ando}(2006)}]{doi:10.1143/JPSJ.75.074716}%
  \BibitemOpen
  \bibfield  {author} {\bibinfo {author} {\bibfnamefont {T.}~\bibnamefont
  {Ando}},\ }\href {\doibase 10.1143/JPSJ.75.074716} {\bibfield  {journal}
  {\bibinfo  {journal} {Journal of the Physical Society of Japan}\ }\textbf
  {\bibinfo {volume} {75}},\ \bibinfo {pages} {074716} (\bibinfo {year}
  {2006})},\ \Eprint
  {http://arxiv.org/abs/http://dx.doi.org/10.1143/JPSJ.75.074716}
  {http://dx.doi.org/10.1143/JPSJ.75.074716} \BibitemShut {NoStop}%
\bibitem [{\citenamefont {Ishigami}\ \emph {et~al.}(2007)\citenamefont
  {Ishigami}, \citenamefont {Chen}, \citenamefont {Cullen}, \citenamefont
  {Fuhrer},\ and\ \citenamefont {Williamsâ}}]{doi:10.1021/nl070613a}%
  \BibitemOpen
  \bibfield  {author} {\bibinfo {author} {\bibfnamefont {M.}~\bibnamefont
  {Ishigami}}, \bibinfo {author} {\bibfnamefont {J.~H.}\ \bibnamefont {Chen}},
  \bibinfo {author} {\bibfnamefont {W.~G.}\ \bibnamefont {Cullen}}, \bibinfo
  {author} {\bibfnamefont {M.~S.}\ \bibnamefont {Fuhrer}}, \ and\ \bibinfo
  {author} {\bibfnamefont {E.~D.}\ \bibnamefont {Williamsâ}},\ }\href {\doibase
  10.1021/nl070613a} {\bibfield  {journal} {\bibinfo  {journal} {Nano Letters}\
  }\textbf {\bibinfo {volume} {7}},\ \bibinfo {pages} {1643} (\bibinfo {year}
  {2007})},\ \bibinfo {note} {pMID: 17497819},\ \Eprint
  {http://arxiv.org/abs/http://dx.doi.org/10.1021/nl070613a}
  {http://dx.doi.org/10.1021/nl070613a} \BibitemShut {NoStop}%
\bibitem [{\citenamefont {Morozov}\ \emph {et~al.}(2008)\citenamefont
  {Morozov}, \citenamefont {Novoselov}, \citenamefont {Katsnelson},
  \citenamefont {Schedin}, \citenamefont {Elias}, \citenamefont {Jaszczak},\
  and\ \citenamefont {Geim}}]{PhysRevLett.100.016602}%
  \BibitemOpen
  \bibfield  {author} {\bibinfo {author} {\bibfnamefont {S.~V.}\ \bibnamefont
  {Morozov}}, \bibinfo {author} {\bibfnamefont {K.~S.}\ \bibnamefont
  {Novoselov}}, \bibinfo {author} {\bibfnamefont {M.~I.}\ \bibnamefont
  {Katsnelson}}, \bibinfo {author} {\bibfnamefont {F.}~\bibnamefont {Schedin}},
  \bibinfo {author} {\bibfnamefont {D.~C.}\ \bibnamefont {Elias}}, \bibinfo
  {author} {\bibfnamefont {J.~A.}\ \bibnamefont {Jaszczak}}, \ and\ \bibinfo
  {author} {\bibfnamefont {A.~K.}\ \bibnamefont {Geim}},\ }\href {\doibase
  10.1103/PhysRevLett.100.016602} {\bibfield  {journal} {\bibinfo  {journal}
  {Phys. Rev. Lett.}\ }\textbf {\bibinfo {volume} {100}},\ \bibinfo {pages}
  {016602} (\bibinfo {year} {2008})}\BibitemShut {NoStop}%
\bibitem [{\citenamefont {Katsnelson}\ and\ \citenamefont
  {Geim}(2008)}]{Katsnelson195}%
  \BibitemOpen
  \bibfield  {author} {\bibinfo {author} {\bibfnamefont {M.}~\bibnamefont
  {Katsnelson}}\ and\ \bibinfo {author} {\bibfnamefont {A.}~\bibnamefont
  {Geim}},\ }\href {\doibase 10.1098/rsta.2007.2157} {\bibfield  {journal}
  {\bibinfo  {journal} {Philosophical Transactions of the Royal Society of
  London A: Mathematical, Physical and Engineering Sciences}\ }\textbf
  {\bibinfo {volume} {366}},\ \bibinfo {pages} {195} (\bibinfo {year}
  {2008})},\ \Eprint
  {http://arxiv.org/abs/http://rsta.royalsocietypublishing.org/content/366/1863/195.full.pdf}
  {http://rsta.royalsocietypublishing.org/content/366/1863/195.full.pdf}
  \BibitemShut {NoStop}%
\bibitem [{\citenamefont {Sevinçli}\ and\ \citenamefont
  {Brandbyge}(2014)}]{AtomicSteps}%
  \BibitemOpen
  \bibfield  {author} {\bibinfo {author} {\bibfnamefont {H.}~\bibnamefont
  {Sevinçli}}\ and\ \bibinfo {author} {\bibfnamefont {M.}~\bibnamefont
  {Brandbyge}},\ }\href {\doibase http://dx.doi.org/10.1063/1.4898066}
  {\bibfield  {journal} {\bibinfo  {journal} {Applied Physics Letters}\
  }\textbf {\bibinfo {volume} {105}},\ \bibinfo {eid} {153108} (\bibinfo {year}
  {2014}),\ http://dx.doi.org/10.1063/1.4898066}\BibitemShut {NoStop}%
\bibitem [{\citenamefont {Qiu}\ and\ \citenamefont {Ruan}(2012)}]{MD_graphene}%
  \BibitemOpen
  \bibfield  {author} {\bibinfo {author} {\bibfnamefont {B.}~\bibnamefont
  {Qiu}}\ and\ \bibinfo {author} {\bibfnamefont {X.}~\bibnamefont {Ruan}},\
  }\href {\doibase http://dx.doi.org/10.1063/1.4712041} {\bibfield  {journal}
  {\bibinfo  {journal} {Applied Physics Letters}\ }\textbf {\bibinfo {volume}
  {100}},\ \bibinfo {eid} {193101} (\bibinfo {year} {2012}),\
  http://dx.doi.org/10.1063/1.4712041}\BibitemShut {NoStop}%
\bibitem [{\citenamefont {Ong}\ \emph {et~al.}(2011)\citenamefont {Ong},
  \citenamefont {Pop},\ and\ \citenamefont {Shiomi}}]{PhysRevB.84.165418}%
  \BibitemOpen
  \bibfield  {author} {\bibinfo {author} {\bibfnamefont {Z.-Y.}\ \bibnamefont
  {Ong}}, \bibinfo {author} {\bibfnamefont {E.}~\bibnamefont {Pop}}, \ and\
  \bibinfo {author} {\bibfnamefont {J.}~\bibnamefont {Shiomi}},\ }\href
  {\doibase 10.1103/PhysRevB.84.165418} {\bibfield  {journal} {\bibinfo
  {journal} {Phys. Rev. B}\ }\textbf {\bibinfo {volume} {84}},\ \bibinfo
  {pages} {165418} (\bibinfo {year} {2011})}\BibitemShut {NoStop}%
\bibitem [{\citenamefont {Fratini}\ and\ \citenamefont
  {Guinea}(2008)}]{PhysRevB.77.195415}%
  \BibitemOpen
  \bibfield  {author} {\bibinfo {author} {\bibfnamefont {S.}~\bibnamefont
  {Fratini}}\ and\ \bibinfo {author} {\bibfnamefont {F.}~\bibnamefont
  {Guinea}},\ }\href {\doibase 10.1103/PhysRevB.77.195415} {\bibfield
  {journal} {\bibinfo  {journal} {Phys. Rev. B}\ }\textbf {\bibinfo {volume}
  {77}},\ \bibinfo {pages} {195415} (\bibinfo {year} {2008})}\BibitemShut
  {NoStop}%
\bibitem [{\citenamefont {Perebeinos}\ and\ \citenamefont
  {Avouris}(2010)}]{PhysRevB.81.195442}%
  \BibitemOpen
  \bibfield  {author} {\bibinfo {author} {\bibfnamefont {V.}~\bibnamefont
  {Perebeinos}}\ and\ \bibinfo {author} {\bibfnamefont {P.}~\bibnamefont
  {Avouris}},\ }\href {\doibase 10.1103/PhysRevB.81.195442} {\bibfield
  {journal} {\bibinfo  {journal} {Phys. Rev. B}\ }\textbf {\bibinfo {volume}
  {81}},\ \bibinfo {pages} {195442} (\bibinfo {year} {2010})}\BibitemShut
  {NoStop}%
\bibitem [{\citenamefont {Chen}\ \emph {et~al.}(2008)\citenamefont {Chen},
  \citenamefont {Jang}, \citenamefont {Xiao}, \citenamefont {Ishigami},\ and\
  \citenamefont {Fuhrer}}]{Chen2008}%
  \BibitemOpen
  \bibfield  {author} {\bibinfo {author} {\bibfnamefont {J.-H.}\ \bibnamefont
  {Chen}}, \bibinfo {author} {\bibfnamefont {C.}~\bibnamefont {Jang}}, \bibinfo
  {author} {\bibfnamefont {S.}~\bibnamefont {Xiao}}, \bibinfo {author}
  {\bibfnamefont {M.}~\bibnamefont {Ishigami}}, \ and\ \bibinfo {author}
  {\bibfnamefont {M.~S.}\ \bibnamefont {Fuhrer}},\ }\href {\doibase
  10.1038/nnano.2008.58} {\bibfield  {journal} {\bibinfo  {journal} {Nat Nano}\
  }\textbf {\bibinfo {volume} {3}},\ \bibinfo {pages} {206} (\bibinfo {year}
  {2008})}\BibitemShut {NoStop}%
\bibitem [{\citenamefont {Bhargavi}\ and\ \citenamefont
  {Kubakaddi}(2013)}]{bhargavi2013scattering}%
  \BibitemOpen
  \bibfield  {author} {\bibinfo {author} {\bibfnamefont {K.}~\bibnamefont
  {Bhargavi}}\ and\ \bibinfo {author} {\bibfnamefont {S.}~\bibnamefont
  {Kubakaddi}},\ }\href@noop {} {\bibfield  {journal} {\bibinfo  {journal}
  {Physica E: Low-dimensional Systems and Nanostructures}\ }\textbf {\bibinfo
  {volume} {52}},\ \bibinfo {pages} {116} (\bibinfo {year} {2013})}\BibitemShut
  {NoStop}%
\bibitem [{\citenamefont {Perebeinos}\ \emph {et~al.}(2009)\citenamefont
  {Perebeinos}, \citenamefont {Rotkin}, \citenamefont {Petrov},\ and\
  \citenamefont {Avouris}}]{doi:10.1021/nl8030086}%
  \BibitemOpen
  \bibfield  {author} {\bibinfo {author} {\bibfnamefont {V.}~\bibnamefont
  {Perebeinos}}, \bibinfo {author} {\bibfnamefont {S.~V.}\ \bibnamefont
  {Rotkin}}, \bibinfo {author} {\bibfnamefont {A.~G.}\ \bibnamefont {Petrov}},
  \ and\ \bibinfo {author} {\bibfnamefont {P.}~\bibnamefont {Avouris}},\ }\href
  {\doibase 10.1021/nl8030086} {\bibfield  {journal} {\bibinfo  {journal} {Nano
  Letters}\ }\textbf {\bibinfo {volume} {9}},\ \bibinfo {pages} {312} (\bibinfo
  {year} {2009})},\ \bibinfo {note} {pMID: 19055370},\ \Eprint
  {http://arxiv.org/abs/http://dx.doi.org/10.1021/nl8030086}
  {http://dx.doi.org/10.1021/nl8030086} \BibitemShut {NoStop}%
\bibitem [{\citenamefont {Petrov}\ and\ \citenamefont {Rotkin}()}]{ref1}%
  \BibitemOpen
  \bibfield  {author} {\bibinfo {author} {\bibfnamefont {A.~G.}\ \bibnamefont
  {Petrov}}\ and\ \bibinfo {author} {\bibfnamefont {S.~V.}\ \bibnamefont
  {Rotkin}},\ }\href {\doibase 10.1134/S0021364006150124} {\bibfield  {journal}
  {\bibinfo  {journal} {JETP Letters}\ }\textbf {\bibinfo {volume} {84}},\
  \bibinfo {pages} {156}}\BibitemShut {NoStop}%
\bibitem [{\citenamefont {Esfarjani}\ \emph {et~al.}(2011)\citenamefont
  {Esfarjani}, \citenamefont {Chen},\ and\ \citenamefont
  {Stokes}}]{esfarjani2011heat}%
  \BibitemOpen
  \bibfield  {author} {\bibinfo {author} {\bibfnamefont {K.}~\bibnamefont
  {Esfarjani}}, \bibinfo {author} {\bibfnamefont {G.}~\bibnamefont {Chen}}, \
  and\ \bibinfo {author} {\bibfnamefont {H.~T.}\ \bibnamefont {Stokes}},\
  }\href@noop {} {\bibfield  {journal} {\bibinfo  {journal} {Physical Review
  B}\ }\textbf {\bibinfo {volume} {84}},\ \bibinfo {pages} {085204} (\bibinfo
  {year} {2011})}\BibitemShut {NoStop}%
\bibitem [{\citenamefont {He}\ \emph {et~al.}(2011)\citenamefont {He},
  \citenamefont {Donadio},\ and\ \citenamefont {Galli}}]{he2011heat}%
  \BibitemOpen
  \bibfield  {author} {\bibinfo {author} {\bibfnamefont {Y.}~\bibnamefont
  {He}}, \bibinfo {author} {\bibfnamefont {D.}~\bibnamefont {Donadio}}, \ and\
  \bibinfo {author} {\bibfnamefont {G.}~\bibnamefont {Galli}},\ }\href@noop {}
  {\bibfield  {journal} {\bibinfo  {journal} {Applied physics letters}\
  }\textbf {\bibinfo {volume} {98}},\ \bibinfo {pages} {144101} (\bibinfo
  {year} {2011})}\BibitemShut {NoStop}%
\bibitem [{\citenamefont {Turney}\ \emph {et~al.}(2009)\citenamefont {Turney},
  \citenamefont {Landry}, \citenamefont {McGaughey},\ and\ \citenamefont
  {Amon}}]{turney2009predicting}%
  \BibitemOpen
  \bibfield  {author} {\bibinfo {author} {\bibfnamefont {J.}~\bibnamefont
  {Turney}}, \bibinfo {author} {\bibfnamefont {E.}~\bibnamefont {Landry}},
  \bibinfo {author} {\bibfnamefont {A.}~\bibnamefont {McGaughey}}, \ and\
  \bibinfo {author} {\bibfnamefont {C.}~\bibnamefont {Amon}},\ }\href@noop {}
  {\bibfield  {journal} {\bibinfo  {journal} {Physical Review B}\ }\textbf
  {\bibinfo {volume} {79}},\ \bibinfo {pages} {064301} (\bibinfo {year}
  {2009})}\BibitemShut {NoStop}%
\bibitem [{\citenamefont {Bao}\ \emph {et~al.}(2012)\citenamefont {Bao},
  \citenamefont {Qiu}, \citenamefont {Zhang},\ and\ \citenamefont
  {Ruan}}]{bao2012first}%
  \BibitemOpen
  \bibfield  {author} {\bibinfo {author} {\bibfnamefont {H.}~\bibnamefont
  {Bao}}, \bibinfo {author} {\bibfnamefont {B.}~\bibnamefont {Qiu}}, \bibinfo
  {author} {\bibfnamefont {Y.}~\bibnamefont {Zhang}}, \ and\ \bibinfo {author}
  {\bibfnamefont {X.}~\bibnamefont {Ruan}},\ }\href@noop {} {\bibfield
  {journal} {\bibinfo  {journal} {Journal of Quantitative Spectroscopy and
  Radiative Transfer}\ }\textbf {\bibinfo {volume} {113}},\ \bibinfo {pages}
  {1683} (\bibinfo {year} {2012})}\BibitemShut {NoStop}%
\bibitem [{\citenamefont {Anand}\ \emph {et~al.}(1996)\citenamefont {Anand},
  \citenamefont {Verma}, \citenamefont {Jain},\ and\ \citenamefont
  {Abbi}}]{anand1996temperature}%
  \BibitemOpen
  \bibfield  {author} {\bibinfo {author} {\bibfnamefont {S.}~\bibnamefont
  {Anand}}, \bibinfo {author} {\bibfnamefont {P.}~\bibnamefont {Verma}},
  \bibinfo {author} {\bibfnamefont {K.}~\bibnamefont {Jain}}, \ and\ \bibinfo
  {author} {\bibfnamefont {S.}~\bibnamefont {Abbi}},\ }\href@noop {} {\bibfield
   {journal} {\bibinfo  {journal} {Physica B: Condensed Matter}\ }\textbf
  {\bibinfo {volume} {226}},\ \bibinfo {pages} {331} (\bibinfo {year}
  {1996})}\BibitemShut {NoStop}%
\bibitem [{\citenamefont {Aku-Leh}\ \emph {et~al.}(2005)\citenamefont
  {Aku-Leh}, \citenamefont {Zhao}, \citenamefont {Merlin}, \citenamefont
  {Menendez},\ and\ \citenamefont {Cardona}}]{aku2005long}%
  \BibitemOpen
  \bibfield  {author} {\bibinfo {author} {\bibfnamefont {C.}~\bibnamefont
  {Aku-Leh}}, \bibinfo {author} {\bibfnamefont {J.}~\bibnamefont {Zhao}},
  \bibinfo {author} {\bibfnamefont {R.}~\bibnamefont {Merlin}}, \bibinfo
  {author} {\bibfnamefont {J.}~\bibnamefont {Menendez}}, \ and\ \bibinfo
  {author} {\bibfnamefont {M.}~\bibnamefont {Cardona}},\ }\href@noop {}
  {\bibfield  {journal} {\bibinfo  {journal} {Physical Review B}\ }\textbf
  {\bibinfo {volume} {71}},\ \bibinfo {pages} {205211} (\bibinfo {year}
  {2005})}\BibitemShut {NoStop}%
\bibitem [{\citenamefont {Song}\ \emph {et~al.}(2008)\citenamefont {Song},
  \citenamefont {Wang}, \citenamefont {Dukovic}, \citenamefont {Zheng},
  \citenamefont {Semke}, \citenamefont {Brus},\ and\ \citenamefont
  {Heinz}}]{song2008direct}%
  \BibitemOpen
  \bibfield  {author} {\bibinfo {author} {\bibfnamefont {D.}~\bibnamefont
  {Song}}, \bibinfo {author} {\bibfnamefont {F.}~\bibnamefont {Wang}}, \bibinfo
  {author} {\bibfnamefont {G.}~\bibnamefont {Dukovic}}, \bibinfo {author}
  {\bibfnamefont {M.}~\bibnamefont {Zheng}}, \bibinfo {author} {\bibfnamefont
  {E.}~\bibnamefont {Semke}}, \bibinfo {author} {\bibfnamefont {L.~E.}\
  \bibnamefont {Brus}}, \ and\ \bibinfo {author} {\bibfnamefont {T.~F.}\
  \bibnamefont {Heinz}},\ }\href@noop {} {\bibfield  {journal} {\bibinfo
  {journal} {Physical review letters}\ }\textbf {\bibinfo {volume} {100}},\
  \bibinfo {pages} {225503} (\bibinfo {year} {2008})}\BibitemShut {NoStop}%
\bibitem [{\citenamefont {Letcher}\ \emph {et~al.}(2007)\citenamefont
  {Letcher}, \citenamefont {Kang}, \citenamefont {Cahill},\ and\ \citenamefont
  {Dlott}}]{letcher2007effects}%
  \BibitemOpen
  \bibfield  {author} {\bibinfo {author} {\bibfnamefont {J.~J.}\ \bibnamefont
  {Letcher}}, \bibinfo {author} {\bibfnamefont {K.}~\bibnamefont {Kang}},
  \bibinfo {author} {\bibfnamefont {D.~G.}\ \bibnamefont {Cahill}}, \ and\
  \bibinfo {author} {\bibfnamefont {D.~D.}\ \bibnamefont {Dlott}},\ }\href@noop
  {} {\bibfield  {journal} {\bibinfo  {journal} {Applied physics letters}\
  }\textbf {\bibinfo {volume} {90}},\ \bibinfo {pages} {252104} (\bibinfo
  {year} {2007})}\BibitemShut {NoStop}%
\bibitem [{\citenamefont {Liu}\ \emph {et~al.}(2000)\citenamefont {Liu},
  \citenamefont {Bursill}, \citenamefont {Prawer},\ and\ \citenamefont
  {Beserman}}]{PhysRevB.61.3391}%
  \BibitemOpen
  \bibfield  {author} {\bibinfo {author} {\bibfnamefont {M.~S.}\ \bibnamefont
  {Liu}}, \bibinfo {author} {\bibfnamefont {L.~A.}\ \bibnamefont {Bursill}},
  \bibinfo {author} {\bibfnamefont {S.}~\bibnamefont {Prawer}}, \ and\ \bibinfo
  {author} {\bibfnamefont {R.}~\bibnamefont {Beserman}},\ }\href {\doibase
  10.1103/PhysRevB.61.3391} {\bibfield  {journal} {\bibinfo  {journal} {Phys.
  Rev. B}\ }\textbf {\bibinfo {volume} {61}},\ \bibinfo {pages} {3391}
  (\bibinfo {year} {2000})}\BibitemShut {NoStop}%
\bibitem [{\citenamefont {Lee}\ \emph {et~al.}(2010)\citenamefont {Lee},
  \citenamefont {Sussman}, \citenamefont {Nunn}, \citenamefont {Lorenz},
  \citenamefont {Reim}, \citenamefont {Jaksch}, \citenamefont {Walmsley},
  \citenamefont {Spizzirri},\ and\ \citenamefont {Prawer}}]{lee2010comparing}%
  \BibitemOpen
  \bibfield  {author} {\bibinfo {author} {\bibfnamefont {K.}~\bibnamefont
  {Lee}}, \bibinfo {author} {\bibfnamefont {B.~J.}\ \bibnamefont {Sussman}},
  \bibinfo {author} {\bibfnamefont {J.}~\bibnamefont {Nunn}}, \bibinfo {author}
  {\bibfnamefont {V.}~\bibnamefont {Lorenz}}, \bibinfo {author} {\bibfnamefont
  {K.}~\bibnamefont {Reim}}, \bibinfo {author} {\bibfnamefont {D.}~\bibnamefont
  {Jaksch}}, \bibinfo {author} {\bibfnamefont {I.}~\bibnamefont {Walmsley}},
  \bibinfo {author} {\bibfnamefont {P.}~\bibnamefont {Spizzirri}}, \ and\
  \bibinfo {author} {\bibfnamefont {S.}~\bibnamefont {Prawer}},\ }\href@noop {}
  {\bibfield  {journal} {\bibinfo  {journal} {Diamond and Related Materials}\
  }\textbf {\bibinfo {volume} {19}},\ \bibinfo {pages} {1289} (\bibinfo {year}
  {2010})}\BibitemShut {NoStop}%
\bibitem [{\citenamefont {Qi}\ \emph {et~al.}(2010)\citenamefont {Qi},
  \citenamefont {Rhim}, \citenamefont {Sun}, \citenamefont {Weinert},\ and\
  \citenamefont {Li}}]{qi2010epitaxial}%
  \BibitemOpen
  \bibfield  {author} {\bibinfo {author} {\bibfnamefont {Y.}~\bibnamefont
  {Qi}}, \bibinfo {author} {\bibfnamefont {S.}~\bibnamefont {Rhim}}, \bibinfo
  {author} {\bibfnamefont {G.}~\bibnamefont {Sun}}, \bibinfo {author}
  {\bibfnamefont {M.}~\bibnamefont {Weinert}}, \ and\ \bibinfo {author}
  {\bibfnamefont {L.}~\bibnamefont {Li}},\ }\href@noop {} {\bibfield  {journal}
  {\bibinfo  {journal} {Physical review letters}\ }\textbf {\bibinfo {volume}
  {105}},\ \bibinfo {pages} {085502} (\bibinfo {year} {2010})}\BibitemShut
  {NoStop}%
\bibitem [{\citenamefont {Mahan}(2009)}]{mahan2009kapitza}%
  \BibitemOpen
  \bibfield  {author} {\bibinfo {author} {\bibfnamefont {G.}~\bibnamefont
  {Mahan}},\ }\href@noop {} {\bibfield  {journal} {\bibinfo  {journal}
  {Physical Review B}\ }\textbf {\bibinfo {volume} {79}},\ \bibinfo {pages}
  {075408} (\bibinfo {year} {2009})}\BibitemShut {NoStop}%
\end{thebibliography}%

\end{document}